# Numerical Calculation of Electric Field Enhancement in Neutron Traps with Rough Walls Coated with Superfluid Helium

V. D. Kochev[a], T. I. Mogilyuk[b], S. S. Kostenko[c], and P. D. Grigoriev[a, d], *

[a] *National University of Science and Technology MISIS, Moscow, 119049 Russia*
[b] *National Research Centre "Kurchatov Institute," Moscow, 123182 Russia*
[c] *Federal Research Center of Problems of Chemical Physics and Medicinal Chemistry, Russian Academy of Sciences, Chernogolovka, Moscow oblast, 142432 Russia*
[d] *Landau Institute for Theoretical Physics, Russian Academy of Sciences, Chernogolovka, Moscow oblast, 142432 Russia*
*e-mail: grigorev@itp.ac.ru

**Abstract**—A film of liquid helium on the surface of material traps for ultracold neutrons protects the neutrons from being absorbed by the trap walls. By using surface roughness and an electrostatic field, it is possible to maintain a helium film of sufficient thickness throughout the height of the trap. The field distribution near the tip of such wall roughness of the trap was calculated, and the effect of this field on holding the helium film was estimated.

## INTRODUCTION

For particle physics, astrophysics, and cosmology, a very important parameter is the neutron lifetime $\tau_n$ [1–5]. The process of primary nucleosynthesis after the Big Bang depends on it, and in combination with data of spin–electron asymmetry measurements in experiments on the decay of polarized neutrons [6–8], one can obtain the ratio of the vector and axial coupling constants of the weak interaction.

To date, precision measurements of the neutron lifetime $\tau_n$ are made using special traps (so-called "neutron bottles" [9–13]) for ultracold neutrons (UCNs), in which UCNs are held on top by a gravitational field; and below and on the sides, by a material that weakly absorbs neutrons and creates a potential barrier $V_0$. Recent measurements using such traps [12] have given the value $\tau_n$ = 881.5 ± 0.7 (stat.) ± 0.6 (syst.) s (here and below, estimates of statistical and systematic errors are also given). The presence of a magnetic moment in a neutron makes it possible to use the method of magneto-gravitational trapping of UCNs [14–19], and recent measurements using this method [19] have given $\tau_n$ = 877.75 ± 0.28 (stat.) +0.22/−0.16 (syst.) s. The stated error of these methods does not exceed 1 s, but the obtained values of $\tau_n$ differ by almost 4 s; i.e., the confidence intervals of the values obtained by different methods do not even overlap. Presumably, this discrepancy arises from an underestimation of UCN losses in both material and magnetic traps. In magnetic traps, this underestimation may be due to neutron spin flip in a nonuniform magnetic field $\vec{B}$, which, due to the condition $\text{div}\vec{B} = 0$ and the heterogeneity of $\vec{B}$ except components along the $z$ axis, necessarily contains perpendicular components that flip the spin. In material traps, errors in estimating UCN losses due to neutron absorption by trap walls can arise due to wall roughness, unaccounted impurities, geometric extrapolation errors, etc. The main alternative to using UCNs for measurement is $\tau_n$ is the cold neutron beam method [20–22], which gives $\tau_n$ = 887.7 ± 1.2 (stat.) ± 1.9 (syst.) s [21], and this value differs even more from the values obtained by other methods, which is a known and still unresolved problem.

The rate of neutron leakage from UCN material traps depends on the type of material with which the walls and bottom of the trap are coated. At the moment, Fomblin oil is most often used [9–12], which is kept at a low temperature ($T$ < 90 K) in order to reduce inelastic neutron scattering. A major concern is surface roughness, which makes it impossible to accurately calculate the probability of UCN loss in each collision. The surface-to-volume ratio in material traps and UCN losses on trap walls can be reduced by increasing the size of the trap; however, further increasing the size of UCN traps seems technically difficult and not very useful, since the main loss of neutrons occurs when they collide with the bottom of the trap, rather than with the walls. The rate of neutron collisions with the bottom of the trap is determined by the kinetic energy of UCNs along the $z$ axis and does not depend on the size of the trap. Therefore, the measurement accuracy $\tau_n$ in conventional UCN traps appears to have reached its limit.

## USING SUPERFLUID HELIUM AS A BARRIER TO NEUTRONS

A possible improvement to the technique, which theoretically allows for a further reduction in the rate of neutron leakage from UCN traps, is our proposed [23–25] coating of the rough walls of the trap due to van der Waals attraction with a thin film of liquid superfluid $^4$He, which does not absorb neutrons [26–28]. The main problem with this idea is that the $^4$He film on the side walls is too thin: $d_{He} \approx 10$ nm, while the penetration depth of neutrons into liquid $^4$He is $\kappa_{He}^{-1} = \hbar/\sqrt{2m_n V_0^{He}} \approx 33.3$ nm. This problem is solved by applying a potential difference between the side wall of the UCN trap and the electrode, which leads to an increase in the electric field near the wall roughness. As a result, $^4$He is attracted to the tips of the roughness, and the thickness of the layer is leveled (see Fig. 1 in [25]). Artificially roughening and applying electrical voltage make sense only to the side walls of the UCN trap, since the thickness of the helium film at the bottom of the trap is determined only by the amount of liquid helium and can be made arbitrary. A regular material UCN trap does not have a top cover; the UCNs on top are held in place by gravity. To avoid loss of helium due to rising up the walls and further leakage from the trap, one can (1) use a lid, (2) pour helium from the other side of the walls, or (3) prevent helium from flowing along the walls of the trap. The applied electric field and the rough surface of the wall can trap helium, since its flow through the tip is now complicated by both the thinning of the film and overcoming the electrostatic attraction to the high-field region. Leakage and other losses of helium from the UCN trap require further study. It is important that the helium remain trapped during the beta decay of the neutron ($\tau_n \approx 15$ min).

The proposed configuration, however, has a number of disadvantages. Firstly, $^4$He provides a very low optical potential barrier $V_0^{He} \approx 18.5$ neV for neutrons, which is significantly less than the barrier $V_0^F \approx 106$ neV of Fomblin oil. The phase volume of UCNs and their density in the $^4$He trap are reduced by a factor of $(V_0^F/V_0^{He})^{3/2} \approx 13.7$ in comparison with the Fomblin coating, which increases the statistical error. However, this disadvantage is compensated by the development of neutron-generators [29, 30]. Secondly, a very low temperature of $T < 0.5$ K is required, at which the con-centration of $^4$He vapor, which inelastically scatters neutrons, is exponentially small (e.g., the evaporation energy of one atom $^4$He is 7.17 K $\approx 0.62$ meV). At higher temperatures, $^4$He vapor inelastically dissipates UCNs, giving them energy $k_B T \gg V_0^{He}$.

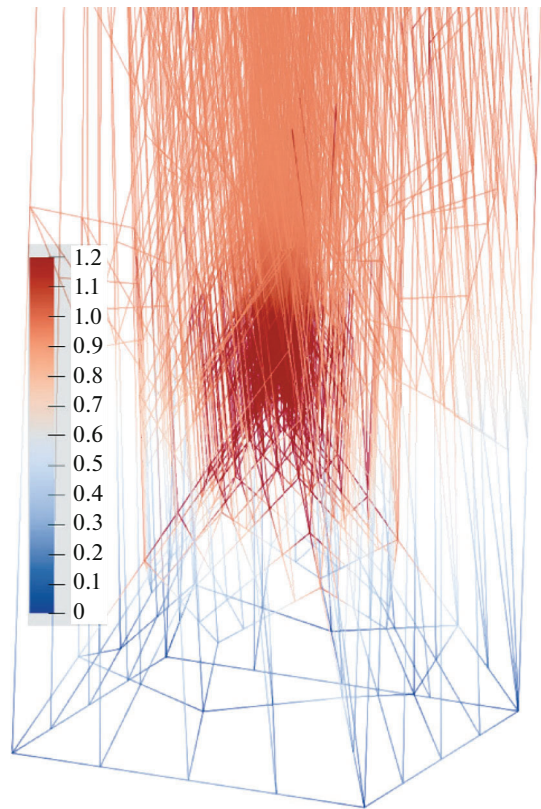

**Fig. 1.** Finite element calculation mesh for pyramids of size $l_R = h_R = 1$ μm. The color shows the distribution of the electric field gain $E(\vec{r})/E_0$.

The main source of inelastic scattering of UCNs in a trap with liquid $^4$He at temperatures below 0.5 K are ripplons—surface wave quanta, leading to a linear dependence of the neutron scattering rate on temperature [31], which persists even at ultralow temperatures. However, the force of interaction between neutrons and ripplons is quite weak, which makes it possible to store UCNs in traps coated with $^4$He. Moreover, the linear dependence of UCN losses on their scattering by ripplons is very convenient for accounting for this systematic error.

In our previous work [25], we proposed variants of the geometry of the electrode in the trap and made estimates of the potential difference between the electrode and the trap that is necessary to achieve sufficient attraction of $^4$He, which is ~0.5–1.2 kV, depending on the selected geometry. In addition, the effect of this electric field on the dispersion of ripplons was assessed. However, numerical calculations and quan-titative estimates of the electrical voltage that is required to hold a liquid $^4$He film of a necessary thickness was obtained [25] only for two-dimensional artificial wall roughness in the form of periodic triangular grooves. In this work, we perform a numerical calculation for a three-dimensional artificial wall roughness in the form of a periodic mesh of square pyramids. It is expected that the 3D model of a rough wall produces a greater electric field enhancement near the tip of the roughness (i.e., the top of the pyramids in our case) than the 2D artificial roughness. This could simplify the creation of new types of UCN traps needed to measure the neutron beta decay time and for other neutron experiments.

The effect of electric field enhancement near the top of a sharp equipotential needle with a radius $r_e$ [32] in comparison with periodic two-dimensional roughness was analyzed in Eqs. (27) and (28) of our paper [25], as well as in the discussion after these equations. The corresponding gain is given by the formula $E(r)/E_0 \simeq l_R(r+r_e/2)^{-1}$, where $l_R$ is is the period of roughness, and $r$ is the distance to the needle tip. Because of the transition from two-dimensional to three-dimensional wall roughness, this gain turns out to be significant and makes it possible to reduce the applied electrical voltage by several times. However, in real experiments, a roughness in the form of thin needles corresponding to a zero apex angle will be too brittle; therefore, it is useful to calculate the electric field enhancement near a three-dimensional roughness in the form of a pyramid with a finite apex angle $\gtrsim 45°$. Such pyramids will be much less fragile and more resistant to both mechanical and electrostatic stress. Below are the results of this calculation and their analysis.

## NUMERICAL CALCULATION OF THE ELECTRIC FIELD ENHANCEMENT NEAR WALL ROUGHNESS

In article [25], we analytically derived formula (13) for the electric field gain $E(r)/E_0$ near the triangular tip of the roughness in the two-dimensional case:

$$\frac{E(r)}{E_0} = \frac{1}{1+2\beta/\pi}\left(\frac{l_R}{r}\right)^{4\beta/(\pi+2\beta)}, \quad (1)$$

where $\beta = \arctan(2h_R/l_R)$ is the angle between the side of the triangular roughness and the width of the roughness $l_R$, and $h_R$ is the roughness height.

In this work, we consider a three-dimensional model of a rough wall in the form of an equipotential, periodic in two directions ($\varphi = 0$), grid of square pyramids located at a distance from the flat electrode ($\varphi = \varphi_0$). The parameters $l_R$ and $h_R$ are specified, respectively, as the width of the base and the height of the pyramid (see Fig. 1 in [25]). The pyramids are closely adjacent to each other; i.e. $l_R$ is also the distance between the vertices of the pyramids and the period along both directions along the side wall.

We numerically solved the Laplace equation for the electric potential $\nabla^2\varphi(\vec{r}) = 0$ with potential difference $\varphi_0$ between a rough mesh and an electrode and under periodic boundary conditions between the pyramids, after which the electric field distribution $\vec{E}(\vec{r}) = -\nabla\varphi(\vec{r})$ in the volume of the problem was found. The equation was solved by the finite element method using the deal.II package [33]. The initial finite element mesh was generated using the Gmsh package [34]; an example of a computational mesh is shown in Fig. 1.

Figure 2 shows the dependence of the electric field gain $E(r)/E_0$ on the distance to the tip of the pyramid. The height of the pyramid was taken equal to $h_R = 1$ μm. It can be seen that, for a sharp pyramid with a narrow base of $l_R = 1$ μm, the field near the tip is much stronger than that for a flat pyramid with a base of $l_R = 10$ μm. For example, at a distance of 0.025 μm, the field is enhanced by a factor of approximately 3 and 1.9, respectively. At a distance of 0.1 μm in both cases, the field is enhanced by a factor of approximately 1.5. The dashed line corresponds to formula (1); it can be seen that the effects in the two-dimensional and three-dimensional cases are not qualitatively different.

Figure 3 presents the same graph in a log–log scale (distance $r$ is normalized to base width $l_R$). It can be seen that, in both cases, the dependence can be fitted by a power law $E(r) \propto r^{-\alpha}$ at $\alpha \approx 0.5$ and 0.2, respectively. This result is consistent with the analysis of the Laplace equation under similar boundary conditions (with a cone instead of a pyramid) [35], but the exponent $\alpha$ has not been previously estimated. Moreover, that analysis [35] considered a unit cone, rather than a periodic system as in our case.

Let us now estimate the magnitude $E_0$ of the electric field far from the roughness that is required to achieve sufficient attraction of $^4$He to roughness. Only neutrons with the kinetic energy $E < V_0^{He}$ can be retained in the trap; this corresponds to the height of neutrons above the bottom of the "neutron bottle" of no more that $h_{max} = V_0^{He}/m_n g \approx 18$ cm. At this height, the electric field (without taking into account the gain, i.e., if we consider this field uniform) should be no less than $E \geq E_* = \sqrt{4\pi\rho_{He}gh/(\varepsilon_{He}-1)} \approx 230$ kV cm$^{-1}$ [24]. We are interested in the thickness of the superfluid $^4$He film that is sufficient to create a potential barrier to neutrons of $d \approx 0.1$ μm [23]; at this barrier, the gain is approximately equal to $E(r)/E_0 \approx 1.5$, from where the field far from roughness is $E_0 \approx 150$ kV cm$^{-1}$.

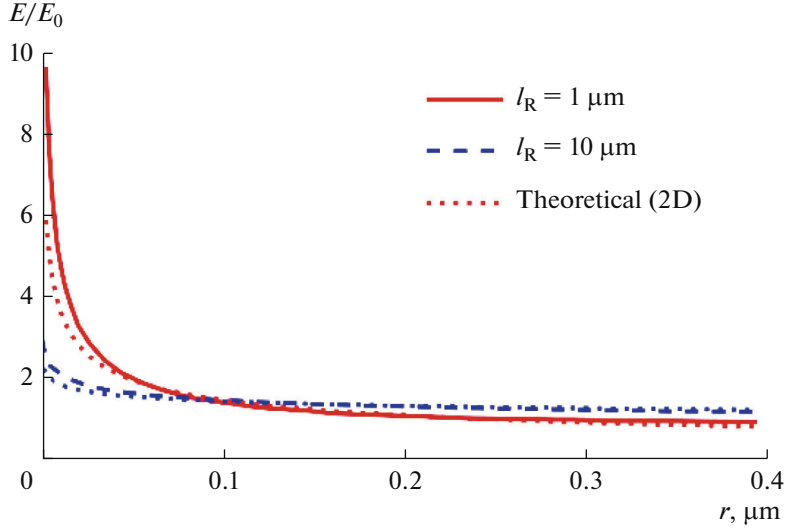

**Fig. 2.** Electric field enhancement near the tip of the roughness of the trap wall.

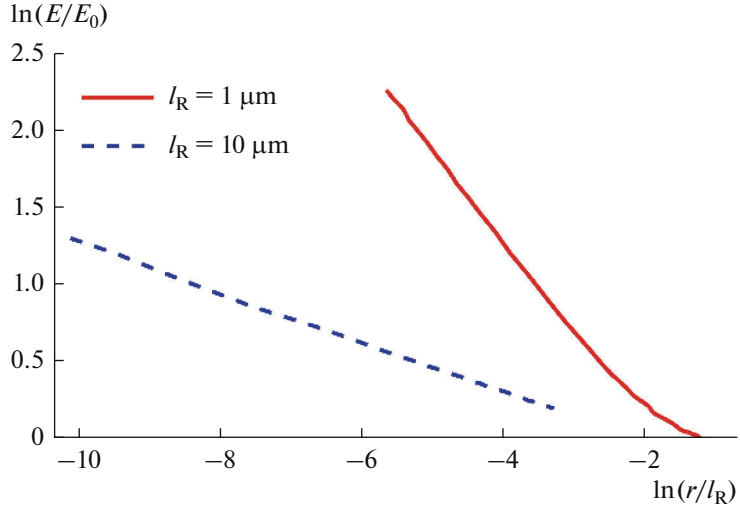

**Fig. 3.** Curves from Fig. 2 in a logarithmic scale.

Although this is a fairly strong field, it is theoretically achievable (an order of magnitude less than the dielectric breakdown field $E_{max} > 1$ MV cm$^{-1}$ [36]).

## DISCUSSION

We investigated how the transition from a two-dimensional artificial roughness of an equipotential surface in the form of periodically spaced triangular grooves to a three-dimensional roughness in the form of periodically spaced square pyramids enhances the electric field near the tops of the pyramids. The two-dimensional roughness corresponds to diffraction gratings, the production of which has long been developed [37] and which are available for purchase. Three-dimensional roughness in the form of quadrangular pyramids is not much more difficult to produce, but still requires expenses; therefore, the transition from two-dimensional to three-dimensional wall roughness is justified only if the gain is noticeable.

Our calculations showed that the gain from such a transition is significant only if the angle of the top of the pyramid is small enough and the distance to the top is not too great. Since the sharp top is more fragile, we calculated the configuration where the side of the pyramid with the base $l_R$ is not less than the height of the pyramid $h_R$, i.e., where the side faces of the pyramid have an angle of inclination of the side face of $\beta \leq 63.4°$, which corresponds to an angle of the vertex of the face of $\theta = \pi - 2\beta \geq 53.2°$. As follows from Figs. 2 and 3, even at this pyramid angle, a significant enhancement of the electric field occurs at a distance of less than 0.1 μm from the top of the pyramid. At a distance greater than 70 nm from the vertex, there is practically no gain due to the transition from two-dimensional to three-dimensional wall roughness. As was shown (see Fig. 4 in [23]), for reliable protection of UCNs from absorption by the vessel wall (to reduce the UCN absorption rate by a factor of more than 100), the thickness of the $^4$He liquid film is required to be more than 0.1 μm. At a thickness of 70 nm, the neutron-energy-averaged rate of UCN losses due to absorption by the vessel walls is only one order of magnitude. This is also significant, since we are talking only about the losses near the top of the roughness, whereas on the main surface area of the wall the $^4$He film thickness exceeds the required 0.1 μm. In addition, to maintain the film of the required thickness at a height less than $h_{\max} \approx 18$ cm (which corresponds to the maximum UCN energy in the trap $E = V_0^{\text{He}} \approx 18.5$ neV) requires a lower electric field. Nevertheless, the gain obtained in our calculations due to the transition from two-dimensional to three-dimensional surface roughness of the UCN trap walls turned out to be less than expected. This gain is probably not worth the effort of creating a 3D pyramid roughness instead of a 2D roughness like the more familiar triangular diffraction grating.

## CONCLUSIONS

The proposed complete coating of the walls of an ultracold neutron (UCN) trap with liquid $^4$He may lead to a new generation of ultracold neutron traps with very long storage times. This may significantly improve the accuracy of neutron lifetime measurements and other ultracold neutron experiments. The work examined the possibility and necessary conditions for the retention of a $^4$He film of the required thickness by surface roughness and electrostatic field. The thinnest helium film occurs near the tip, where the gain from the van der Waals attraction of helium to the walls is smallest. It is near the tip that the electric field on the equipotential surface is maximum and can compensate for the decrease in van der Waals forces. Our numerical calculations of the electric field distribution at various roughness tip shapes showed that the roughness shape in the form of a triangular diffraction grating is almost as effective as the more complex three-dimensional roughness in the form of periodically arranged pyramids.

## FUNDING

This work was supported by the Russian Science Foundation (grant no. 23-22-00312, https://rscf.ru/project/23-22-00312/).

## CONFLICT OF INTEREST

The authors of this work declare that they have no conflicts of interest.

## REFERENCES

1. Abele, H., *Prog. Part. Nucl. Phys.*, 2008, vol. 60, no. 1, p. 1.
2. Ramsey-Musolf, M.J. and Su, S., *Phys. Rep.*, 2008, vol. 456, no. 1, p. 1.
3. Dubbers, D. and Schmidt, M.G., *Rev. Mod. Phys.*, 2011, vol. 83, no. 4, p. 1111.
4. Wietfeldt, F.E. and Greene, G.L., *Rev. Mod. Phys.*, 2011, vol. 83, no. 4, p. 1173.
5. González-Alonso, M., Naviliat-Cuncic, O., and Severijns, N., *Prog. Part. Nucl. Phys.*, 2019, vol. 104, p. 165.
6. Liu, J., Mendenhall, M.P., Holley, A.T., et al., *Phys. Rev. Lett.*, 2010, vol. 105, no. 18, p. 181803.
7. Märkisch, B., Mest, H. Saul, H., et al., *Phys. Rev. Lett.*, 2019, vol. 122, no. 24, p. 242501.
8. Sun, X., Adamek, E., Allgeier, B., et al., *Phys. Rev. C*, 2020, vol. 101, no. 3, p. 035503.
9. Serebrov, A.P., Varlamov, V.E., Kharitonov, A.G., et al., *Phys. Rev. C*, 2008, vol. 78, no. 3, p. 035505.
10. Arzumanov, S., Bondarenko, L., Chernyavsky, S., et al., *Phys. Lett. B*, 2015, vol. 745, p. 79.
11. Serebrov, A.P., Kolomenskiy, E.A., Fomin, A.K., et al., *JETP Lett.*, 2017, vol. 106, no. 10, p. 623.
12. Serebrov, A.P., Kolomenskiy, E.A., Fomin, A.K., et al., *Phys. Rev. C*, 2018, vol. 97, no. 5, p. 055503.
13. Pattie, R., Callahan, N.B., Cude-Woods, C., et al., *EPJ Web Conf.*, 2019, vol. 219, p. 03004.
14. Huffman, P.R., Brome, C.R., Butterworth, J.S., et al., *Nature*, 2000, vol. 403, no. 6765, p. 62.
15. Leung, K.K.H., Geltenbort, P., Ivanov, S., et al., *Phys. Rev. C*, 2016, vol. 94, no. 4, p. 045502.
16. Steyerl, A., Leung, K.K.H., Kaufman, C., et al., *Phys. Rev. C*, 2017, vol. 95, no. 3, p. 035502.
17. Ezhov, V.F., Andreev, A.Z., Bazarov, B.A., et al., *JETP Lett.*, 2018, vol. 107, no. 11, p. 671.
18. Pattie, R.W., Callahan, N.B., Cude-Woods, C., et al., *Science*, 2018, vol. 360, no. 6389, p. 627.
19. Gonzalez, F.M., Fries, E.M., Cude-Woods, C., et al., *Phys. Rev. Lett.*, 2021, vol. 127, no. 16, p. 162501.
20. Nico, J.S., Dewey, M.S., Gilliam, D.M., et al., *Phys. Rev. C*, 2005, vol. 71, no. 5, p. 055502.
21. Yue, A.T., Dewey, M.S., Gilliam, D.M., et al., *Phys. Rev. Lett.*, 2013, vol. 111, no. 22, p. 222501.
22. Hirota, K., Ichikawa, G., and Ieki, S., *Prog. Theor. Exp. Phys.*, 2020, vol. 2020, no. 12, p. 123C02.
23. Grigoriev, P.D. and Dyugaev, A.M., *Phys. Rev. C*, 2021, vol. 104, no. 5, p. 055501.
24. Grigoriev, P.D., Dyugaev, A.M., Mogilyuk, T.I., and Grigoriev, A.D., *JETP Lett.*, 2021, vol. 114, no. 8, p. 493.
25. Grigoriev, P.D., Sadovnikov, A.V., Kochev, V.D., and Dyugaev, A.M., *Phys. Rev. C*, 2023, vol. 108, no. 2, p. 025501.

26. Golub, R., Jewell, C., Ageron, P., et al., *Z. Phys. B*, 1983, vol. 51. no. 3, p. 187.
27. Bokun, R.C., *Sov. J. Nucl. Phys.*, 1984, vol. 40, no. 1, p. 287.
28. Alfimenkov, V.P., Ignatovich, V.K., Mezhov-Deglin, L.P., et al., *Preprint of the Joint Institute for High Temperatures*, Dubna, 2009, no. 3-2009-197.
29. Alekseev, I.E., Belov, S.E., and Ershov, K.V., *Bull. Russ. Acad. Sci.: Phys.*, 2022, vol. 86, no. 9, p. 1088.
30. Grigoriev, S.V., Kovalenko, N.A., Pavlov, K.A., et al., *Bull. Russ. Acad. Sci.: Phys.*, 2023, vol. 87, no. 11, p. 1561.
31. Grigoriev, P.D., Zimmer, O., Grigoriev, A.D., and Ziman, T., *Phys. Rev. C*, 2016, vol. 94, no. 2, p. 025504.
32. Florkowska, B. and Wlodek, R., *IEEE Trans. Electr. Insul.*, 1993, vol. 28, no. 6, p. 932.
33. Arndt, D., Bangerth, W., Davydov, D., et al., *J. Comput. Math. Appl.*, 2021, vol. 81, p. 407.
34. Geuzaine, C. and Remacle, J.-F., *Int. J. Numer. Methods Eng.*, 2009, vol. 79, no. 11, p. 1309.
35. Marchetti, S. and Rozzi, T., *IEEE Trans. Antennas Propag.*, 1990, vol. 38, no. 9, p. 1333.
36. Ito, T.M., Ramsey, J.C., Yao, W., et al., *Rev. Sci. Instrum.*, 2016, vol. 87, no. 4, p. 045113.
37. Bourgin, Y., Jourlin, Y., Parriaux, O., et al., *Opt. Express*, 2010, vol. 18, no. 10, p. 10557.